\documentclass[apj]{emulateapj}

\usepackage{color}
\usepackage{epsf}
\usepackage{ulem}
\usepackage{graphicx}

\newcommand{\scb}{}
\newcommand{\scr}{}
\newcommand{\Om}{\Omega_m}
\newcommand{\Ob}{\Omega_b}
\newcommand{\LCDM}{\rm {\Lambda CDM}}
\newcommand{\OL}{\Omega_\Lambda}
\newcommand{\Ok}{\Omega_K}
\newcommand{\DA}{D\!_A(z)}

\newcommand{\Oh}{\Omega_m h^2}
\newcommand{\Obhh}{\Omega_b h^2}

\newcommand{\hMpc}{h^{-1}{\rm\;Mpc}}
\newcommand{\hGpc}{h^{-1}{\rm\;Gpc}}
\newcommand{\trihGpc}{h^{-3}{\rm\;Gpc^3}}

\newcommand{\Mpc}{{\rm\;Mpc}}

\newcommand{\al}{\alpha}

\newcommand{\Pm}{P_m}
\newcommand{\sig}{\sigma}

\newcommand{\Nb}{{\it N}-body}
\newcommand{\Signl}{\Sigma_{m}}

\newcommand{\Plin}{P_{\rm lin}}

 \newcommand{\Psm}{P_{\rm nw}}

\newcommand{\Clmod}{C_{\rm m, z_i}(\ell)}
\newcommand{\lobs}{\ell_{\rm obs}}
\newcommand{\lfid}{\ell_{\rm fid}}
\newcommand{\DArs}{D_A(z)/r_s}
\newcommand{\DArsf}{[D_A(z)/r_{s}]_{\rm fid}}
\newcommand{\Daf}{D_{A, \rm fid}(z)}

\newcommand{\LRGs}{CMASS1}
\newcommand{\LRGe}{CMASS2}
\newcommand{\LRGn}{CMASS3}
\newcommand{\LRGt}{CMASS4}
\newcommand{\afit}[3]{#1^{+#2}_{-#3}}

\newcommand{\fc}[1]{\ref{fig:#1}}

\newcommand{\DV}{D_V}
\newcommand{\Clobi}[1]{C_{\rm obs,z_i}{#1}}
\newcommand{\Clmoi}[1]{C_{\rm m,z_i}{#1}}

\newcommand{\llan}{\left\langle}
\newcommand{\rran}{\right\rangle}

\newcommand{\vmC}{\mathbb{C}}

\newcommand{\na}{New Astronomy}
\newcommand{\jcap}{Journal of Cosmology and Astroparticle Physics}

\begin{document}
\title{Acoustic scale from the angular power spectra of SDSS-III DR8 photometric luminous galaxies}
\author{
Hee-Jong Seo\altaffilmark{1}, Shirley Ho\altaffilmark{2}, Martin White\altaffilmark{2,3,4}, Antonio J. Cuesta\altaffilmark{5}, Ashley J. Ross\altaffilmark{6}, Shun Saito\altaffilmark{4}, Beth Reid\altaffilmark{2,7}, Nikhil Padmanabhan\altaffilmark{5}, Will J. Percival\altaffilmark{6}, Roland de Putter\altaffilmark{8}, David J. Schlegel\altaffilmark{2}, Daniel J. Eisenstein\altaffilmark{9}, Xiaoying Xu\altaffilmark{13}, Donald P. Schneider\altaffilmark{18,22},  Ramin Skibba\altaffilmark{13}, Licia Verde\altaffilmark{14,15}, Robert C. Nichol\altaffilmark{6,21}, Dmitry Bizyaev\altaffilmark{20}, Howard Brewington\altaffilmark{20}, J. Brinkmann\altaffilmark{20}, Luiz Alberto Nicolaci da Costa\altaffilmark{10,12}, J. Richard Gott III\altaffilmark{16}, Elena Malanushenko\altaffilmark{20}, Viktor Malanushenko\altaffilmark{20}, Dan Oravetz\altaffilmark{20}, Nathalie Palanque-Delabrouille\altaffilmark{23}, Kaike Pan\altaffilmark{20}, Francisco Prada\altaffilmark{11}, Nicholas P. Ross\altaffilmark{2}, Audrey Simmons\altaffilmark{20}, Fernando de Simoni\altaffilmark{10,12,19}, Alaina Shelden\altaffilmark{20}, Stephanie Snedden\altaffilmark{20}, Idit Zehavi\altaffilmark{17}
}

\begin{abstract}
We measure the acoustic scale from the angular power spectra of the Sloan Digital Sky Survey III (SDSS-III) Data Release 8 imaging catalog that includes $872,921$ galaxies over $\sim 10,000 {\rm deg}^2$ between $0.45<z<0.65$. The extensive spectroscopic training set of the Baryon Oscillation Spectroscopic Survey (BOSS) luminous galaxies allows precise estimates of the true redshift distributions of galaxies in our imaging catalog. Utilizing the redshift distribution information, we build templates and \scr{fit to the power spectra of the data, which are measured in our companion paper, \citet{Ho11},} to derive the location of Baryon acoustic oscillations (BAO) while marginalizing over many free parameters to exclude nearly all of the non-BAO signal. We derive the ratio of the angular diameter distance to the sound horizon scale $\DArs= \afit{9.212}{0.416}{0.404}$ at $z=0.54$, and therefore, $\DA= 1411\pm 65 \Mpc$ at $z=0.54$; the result is fairly independent of assumptions on the underlying cosmology. Our measurement of angular diameter distance $\DA$ is $1.4 \sigma$ higher than what is expected for the concordance $\LCDM$ \citep{Komatsu11}, in accordance to the trend of other spectroscopic BAO measurements for $z\gtrsim 0.35$. We report constraints on cosmological parameters from our measurement in combination with the WMAP7 data and the previous spectroscopic BAO measurements of SDSS \citep{Percival10} and WiggleZ \citep{Blake11b}. We refer to our companion papers \citep{Ho11,dePutter11} for investigations on information of the full power spectrum.
\end{abstract}

\altaffiltext{1}{Berkeley Center for Cosmological Physics, LBL and Department of Physics, University of California, Berkeley, CA 94720, USA}
\altaffiltext{2}{Lawrence Berkeley National Laboratory, 1 Cyclotron Road, Berkeley, CA 94720, USA}
\altaffiltext{3}{Department of Physics, University of California, 366 LeConte Hall, Berkeley, CA 94720, USA}
\altaffiltext{4}{Department of Astronomy, 601 Campbell Hall, University of California at Berkeley, Berkeley, CA 94720, USA}
\altaffiltext{5}{Yale Center for Astronomy and Astrophysics, Yale University, New Haven, CT 06511, USA}
\altaffiltext{6}{Institute of Cosmology \& Gravitation, Dennis Sciama Building, University of Portsmouth, Portsmouth PO1 3FX}
\altaffiltext{7}{Hubble fellow}
\altaffiltext{8}{Instituto de Fisica Corpuscular, Valencia, Spain}
\altaffiltext{9}{Harvard-Smithsonian Center for Astrophysics, 60 Garden St., MS \#20, Cambridge, MA 02138}
\altaffiltext{10}{Laborat\'{o}rio Interinstitucional de e-Astronomia, - LIneA, Rua Gal. Jos\'{e} Cristino 77, Rio de Janeiro, RJ - 20921-400, Brazil}
\altaffiltext{11}{Instituto de Astrofisica de Andalucia (CSIC), E-18008, Granada, Spain}
\altaffiltext{12}{Observat\'{o}rio Nacional, Rua Gal. Jos\'{e} Cristino 77, Rio de Janeiro, RJ 20921-400, Brazil}
\altaffiltext{13}{Steward Observatory, University of Arizona, 933 North Cherry Avenue, Tucson, AZ 85721, USA}
\altaffiltext{14}{Instituci\`{o} Catalana de Recerca i Estudis Avan\c{c}ats, Barcelona, Spain}
\altaffiltext{15}{Institut de Ci\'{e}ncies del Cosmos, Universitat de Barcelona/IEEC, Barcelona 08028, Spain}
\altaffiltext{16}{Department of Astrophysical Sciences, Princeton University, Princeton, NJ 08544, USA}
\altaffiltext{17}{Department of Astronomy, Case Western Reserve University, Cleveland, OH 44106, USA}
\altaffiltext{18}{Institute for Gravitation and the Cosmos, The Pennsylvania State University, University Park, PA 16802}
\altaffiltext{19}{Departamento de F\'isica e Matem\'atica, PURO/Universidade Federal Fluminense, Rua Recife s/n, Jardim Bela Vista, Rio das Ostras, RJ 28890-000, Brasil}
\altaffiltext{20}{Apache Point Observatory, 2001 Apache Point Road, Sunspot, NM 88349, USA}
\altaffiltext{21}{SEPnet, South East Physics Network (www.sepnet.ac.uk)}
\altaffiltext{22}{Department of Astronomy and Astrophysics, The Pennsylvania State University, University Park, PA 16802}
\altaffiltext{23}{CEA, Centre de Saclay, IRFU, 91191 Gif-sur-Yvette, France}

\section{Introduction}
Baryon acoustic oscillations (BAO) imprint a distinct feature in the clustering of photons (i.e., cosmic microwave background), mass, and galaxies. Sound waves that propagated through the hot plasma of photons and baryons in early Universe freeze out as photons and baryons decouple and leave a characteristic oscillatory feature in Fourier space and a single distinct peak in the correlation function approximately\footnote{The scale observed in the mass is not exactly the distance travelled when recombination occurs as the momentum of the baryonic material means that the motion continues for a short time after recombination, until an epoch known as the baryon-drag epoch.} at the distance the sound waves have traveled before the epoch of recombination. The distance is called the ``sound horizon scale'' and determines the physical location of the BAO feature in clustering statistics \citep[e.g.,][]{Peebles70,SZ70,Bond84,Holtzman89,HS96,Hu96,EH98}. 

Cosmic microwave background (CMB) data provides an independent and precise determination of the sound horizon scale. Therefore, comparing this sound horizon scale to the observed location of the BAO from galaxy clustering statistics allows one to constrain the angular diameter distance and Hubble parameters, thereby providing information on the nature of dark energy. This approach is known as the `standard ruler test' \citep[e.g.,][]{Hu96,Eisen03,Blake03,Linder03,Hu03,SE03}.  BAO technique is considered an especially robust dark energy probe \citep{DETF} for various reasons. First, its physical scale is separately measured from CMB data. Second, the nonlinear effects in the matter density field are still mild at the BAO scale $(\sim 150\Mpc)$ such that the resulting systematic effects are small and can be modeled with low-order perturbation theories \citep[e.g.,][]{Meiksin99,SE05,Jeong06,Crocce06b,ESW07,Nishimichi07,Crocce08,Mat08,Pad09,SSEW08,Taruya09,Seo10}. Third, the observational/astrophysical effects such as galaxy/halo bias and redshift distortions are likely smooth in wavenumber and do not mimic BAO such that they can be marginalized over \citep[e.g.,][]{SE05,Huff07,Sanchez08,Pad09,Mehta11} [but see \citet{Dalal11} and \citet{Yoo11} for a possibility of an exotic galaxy bias effect].

In recent years, BAO have been detected in the galaxy distribution and used to constrain cosmology \citep[][]{Eisen05,Cole05,Hutsi06,Tegmark06,Percival07a,Percival07b,Pad07,Blake07,Okumura08,Estra08,Gazt09a,Gazt09b,Percival10,Kazin10,6dF,Crocce11,Blake11a,Blake11b}. Most of these studies have used a 3D distribution of galaxies from spectroscopic surveys to constrain an isotropic distance scale $\DV(z)$ ($\DV(z)\equiv [(1+z)^2D^2_A(z) cz/H(z)]^{1/3}$ where $D_A$ is the angular diameter distance and $H$ is the Hubble parameter) using spherically averaged clustering statistics, while others have constrained $D_A(z)$ and $H(z)$ separately, using anisotropic clustering information. 

Retrieving 3D spatial information requires accurate redshift determination (i.e., spectroscopic surveys), demanding specialized spectrographs and surveys that typically take longer times. Multiband imaging surveys, on the other hand, can more quickly cover a large number of galaxies (low shot noise) and a large area of sky but provide only 2D spatial information, assuming a realistic level of photometric redshift error, and therefore fail to retrieve information on $H(z)$\footnote{A fractional redshift error of 0.25\% in $1+z$ is at least required to recover $H(z)$ \citep{SE03}.}. Another disadvantage of using the imaging data to make BAO measurements is an additional damping of the BAO due to projection effects and difficulty in applying BAO reconstruction. Nevertheless, imaging surveys can, in principle, provide larger and deeper surveys \citep{SE03,Amen05,Blake05,Dolney06,Zhan06} and this prospect has motivated current and future imaging BAO surveys such as the Dark Energy Survey\footnote{www.darkenergysurvey.com} \citep[DES;][]{DES}, the Panoramic Survey Telescope and Rapid Response System\footnote{www.pan-starrs.ifa.hawaii.edu} \citep[PanSTARRS;][]{Pan}, the Physics of the Accelerating Universe survey\footnote{www.ice.cat/pau} \citep[PAU;][]{PAU}, the Large Synoptic Survey Telescope\footnote{www.lsst.org} \citep[LSST;][]{LSST}, and EUCLID\footnote{www.sci.esa.int/euclid} \citep{EUCLID}. 

A number of previous works have analyzed and reported the cosmological constraints from the galaxy clustering of the imaging surveys \citep[e.g.,][]{Tegmark02,Blake07,Ross08,Sawang09,Thomas10,Thomas11,Crocce11,Ross11}, but there have been only a few published works on the BAO measurement \citep{Pad07,Carnero11}. Our goal in this paper is to design a robust method for measuring the location of BAO in the angular power spectrum of imaging surveys and apply it to the final imaging data set of the Sloan Digital Sky Survey III \citep[SDSS;][]{York00}.
 
We use the DR8 imaging catalog of SDSS-III that includes photometric redshifts of luminous galaxies (hereafter, `LGs') between $0.45<z<0.65$ over $\sim 10,000 {\rm deg}^2$ \citep{Ross11}; the spectroscopy from the SDSS-III Baryon Oscillation Spectroscopic Survey \citep[BOSS;][]{Eisen11} is used to create a training sample and therefore to estimate the true redshift distribution of photometric galaxies. The angular power spectra are generated from the data using an optimal quadratic estimator, as presented in detail in \scr{one of our companion papers}, \citet{Ho11}. Utilizing the estimated true redshift distribution, we construct a theoretical BAO template for the angular power spectrum, fit to the location of the observed BAO feature in the angular power spectrum, and derive the angular diameter distance to $z=0.54$ in a manner independent of underlying dark energy models.

The angular clustering of galaxies contains more cosmological information than the scale of BAO. Redshift distortions on very large scales and the overall shape of the power spectrum (e.g., the matter-radiation equality feature) can provide additional information. However, in this paper, we take a very conservative approach and use only the most robust probe, the location of BAO, while excluding most of the non-BAO information. \citet[][Paper I]{Ho11} presents a more extensive study: it includes information of the full power spectrum and derive cosmological constraints. \scr{In parallel, another companion paper, \citet{dePutter11}, measures the mass bound on the sum of neutrino masses using the same power spectra.}

This paper is organized as follows. In \S~\ref{sec:data}, we briefly summarize the imaging data. In \S~\ref{sec:power}, we summarize the method used in Paper I to generate angular power spectra. In \S~\ref{sec:method}, we describe details of the BAO fitting method used in this paper. In \S~\ref{sec:testassume}, we test our method and assumptions using mock data.  In \S~\ref{sec:results}, we present an analysis of the DR8 imaging data and report the best fit angular diameter distance to sound horizon ratio at $z=0.54$. We show the robustness of our result and \scr{show the effect of correcting for the observational systematics}. In \S~\ref{sec:discussions}, we discuss constraints on various cosmological parameters when our BAO measurement is combined with the WMAP7 data and other BAO measurements. Finally, in \S~\ref{sec:con}, we summarize the results in this paper.

\begin{figure}[t]
\plotone{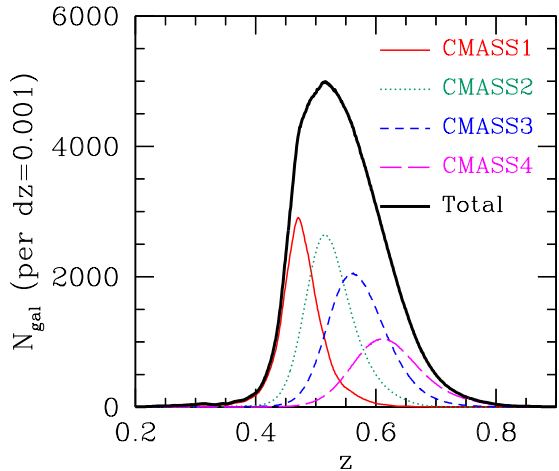}
\caption{The true redshift distribution estimated for our photometric redshift galaxies for the four redshift bins: \LRGs\ for $0.45<z_{\rm ph}<0.5$, \LRGe\ for $0.5<z_{\rm ph}<0.55$, \LRGn\ for $0.55<z_{\rm ph}<0.6$, and \LRGt\ for $0.6<z_{\rm ph}<0.65$. The median and the mean of the combined galaxy distribution is 0.541 and 0.544, respectively.}\label{fig:dndz}
\end{figure}

\section{Data}\label{sec:data}
We use the imaging data of the eighth and final data release \citep[DR8;][]{Aihara11} of SDSS-III \citep[][]{York00} that is obtained by wide-field CCD photometry in five passbands ($u$, $g$, $r$, $i$, $z$) \citep[see][for more technical and data realease details]{Fuku96,Gunn98,Gunn06,Pier03}. We use the photometric redshift catalog constructed as described in \citet{Ross11}\footnote{Available at \url{http://portal.nersc.gov/project/boss/galaxy/photoz}.}. This catalog is selected from DR8 using the same criteria as the SDSS-III BOSS \citep[][]{Eisen11} targets selected to have approximately constant stellar mass \citep[CMASS;][]{White11}. Photometric redshifts and probabilities that an object is a galaxy were obtained using a training sample of 112,778 BOSS CMASS spectra [to be released with Data Release 9 in July 2012] \footnote{We use the redshifts available through MJD 55510}. The final catalog covers $9,913\;{\rm deg}^2$ of sky and consists of $872,921$ galaxies between $0.45<z<0.65$, which is an improvement in the survey area compared to the MegaZ-LRG DR7 catalog \citep[][723,556 objects over $7,746\;{\rm deg}^2$ for $0.45<z<0.65$]{Thomas11}. 

The estimated photometric error, $\sigma_{z_{\rm ph}}$, increases from 0.04 to 0.06 over the redshift range \citep[See Figure 10 of ][]{Ross11}. We define four photometric redshift bins, with widths similar to the photometric error, referred to as \LRGs, \LRGe, \LRGn, \LRGt. Table \ref{tab:tdndz} and Figure \ref{fig:dndz} show the distribution of the effective\footnote{We weight each object by the probability that an object is a galaxy.} number of galaxies in each redshift bins. \scr{Due to the extensive training sample, our determination of the redshift distribution is expected to be quite accurate; for example, based on the Jack-knife resampling of the training sample, we estimate the error on the mean/median of the distribution of each redshift bin to be less than 0.5\%.} The median and mean of the combined galaxy distribution are 0.541 and 0.544, respectively. 

\begin{deluxetable}{c|ccccc}
\tablewidth{0pt}
\tabletypesize{\small}
\tablecaption{\label{tab:tdndz} The four photometric redshift bins.}
\startdata \hline\hline
bins & $z_{\rm ph}$ range & $N_{\rm gal} $ & $\sigma_{z_{\rm ph}}$ & $z_{\rm median}$ & $z_{\rm mean}$\\ \hline
\LRGs\ & 0.45-0.50 &214,971 & 0.043 & 0.474 & 0.475  \\ 
\LRGe\ & 0.50-0.55 & 258,736 & 0.044 & 0.523 & 0.526 \\
\LRGn\ & 0.55-0.60 &248,895  & 0.052 & 0.568 & 0.572\\
\LRGt\ & 0.60-0.65 & 150,319 & 0.063  & 0.617 & 0.621\\
Total  & 0.45-0.65 &  872,921 &        & 0.541& 0.544  
\enddata 
\tablecomments{$N_{\rm gal}$ is the effective number of galaxies after weighting each object by $p_{\rm sg}$, which is the probability that an object is a galaxy. The value of $\sigma_{z_{\rm ph}}$ is the dispersion in redshift for each photo-z bin. }
\end{deluxetable}

\section{Optimal quadratic estimator of angular Power spectra}\label{sec:power}
\begin{figure}[t]
\centerline{\epsfxsize=3in\epsffile{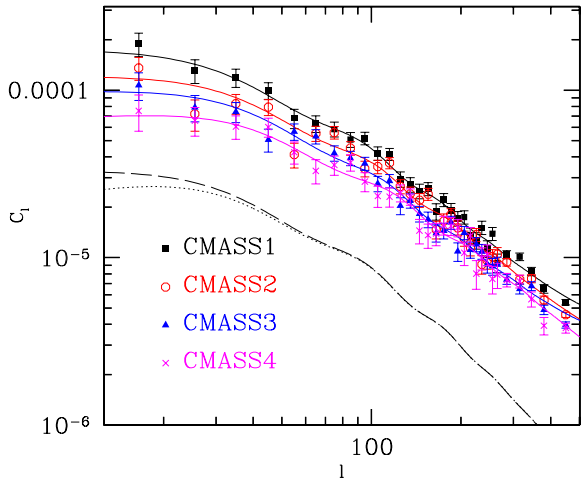}}
\caption{The measured angular power spectra for the four redshift bins. The solid lines show the best fits derived in \S~\ref{sec:results}. \scr{The dashed line is the template for \LRGe\, which is rescaled for clarity, with redshift distortions assuming a galaxy bias of 2. The dotted line is the template without redshift distortions.}}\label{fig:Pk}
\end{figure}

The auto and cross angular power spectra of the four redshift bins were generated using the optimal quadratic estimator in Paper I, to which we refer the readers for more details of the optimal quadratic estimator \citep[OQE, also see][]{Seljak98,Tegmark98,Pad03,Pad07}. 
To summarize, we parametrize the power spectrum with 35 step-function band powers and write the data covariance matrix as  
\begin{eqnarray}
\vmC_{ij} \equiv \llan \delta_i \delta_j \rran =  \sum_\beta p_\beta \vmC_{ij}^{(\beta)} + N_{i}\delta^i_{j},
\end{eqnarray} 
where $\delta_i$ is the galaxy overdensity at the $i^{\rm th}$ pixel at a given redshift bin, $p_\beta$ is the band power for a wave number bin $\beta$, $\vmC^{(\beta)}$ is the derivative of $\vmC$ with respect to the band power $p_\beta$, and $N$ is the shot noise contribution to the covariance matrix, while $\delta^i_{j}$ is the Kronecker delta function. Assuming $\delta_i$ is Gaussian-distributed, requiring the estimator to be unbiased and to have a minimum variance, we derive
 a band power estimator
\begin{eqnarray}
p_\beta = F_{\beta \gamma}^{-1} \left[\frac{1}{2}{\bf\delta}^t \vmC^{-1}\vmC^{(\gamma)}\vmC^{-1} {\bf\delta} -b_{\gamma} \right],
\end{eqnarray} 
where $b_{\gamma}$ is the contribution from $N$ (i.e., $\frac{1}{2}tr[\vmC^{-1}\vmC^{(\gamma)}\vmC^{-1}N]$), i.e., the shot noise, and $F$ is the Fisher information matrix: 
\begin{eqnarray}
F_{\beta \gamma}=\frac{1}{2} tr[\vmC^{-1}\vmC^{(\beta)}\vmC^{-1}\vmC^{(\gamma)}].
\end{eqnarray} 
The variance of the band power is derived by
\begin{eqnarray}
Cov[p_{\beta}, p_{\gamma}]=F^{-1}_{\beta \gamma}.\label{eq:Covab}
\end{eqnarray}
The expected value of band power, $\llan p_{\beta} \rran$, is related to the power spectrum $p(\ell')$ at an integer $\ell'$ by the band window function $W_{\beta \ell'}$ \citep[e.g.,][]{Knox99}:
\begin{eqnarray}
\llan p_{\beta} \rran =\frac{d\llan p_{\beta} \rran }{d p(\ell')}p(\ell')= W_{\beta \ell'} p(\ell'),
\end{eqnarray} 
and 
\begin{eqnarray}
W_{\beta \ell'}=F_{\beta \gamma}^{-1} \frac{1}{2}tr[\vmC^{-1}\vmC^{(\gamma)}\vmC^{-1}\vmC^{(\ell')}], \label{eq:Win}
\end{eqnarray} 
where $\vmC^{(\ell')}$ is the derivative of $\vmC$ with respect to the power at an integer wavenumber $\ell'$.
In this paper, we use the band power $p_\beta$ and the covariance matrix of the band power $F^{-1}$ derived in Paper I. The auto power spectra are shown in Figure \ref{fig:Pk}. The cross-power estimates between different redshift bins were generated in a similar manner and used only for determining covariance between different redshift bins.

\section{Methods}\label{sec:method}
\subsection{Outlines of the fitting method}
Our goal is to robustly measure the location of the BAO scale, and therefore the distance scale $\DA$, while minimizing possible effects from the assumptions made on the nonlinear or/and observational effects as well as cosmology during the fitting procedure. For example, it is non-trivial to properly model the evolution of the broadband power on the nonlinear scales, whether it is due to structure growth, redshift distortions, or galaxy bias. Therefore, we aim to exclude as much non-BAO signal as possible by marginalizing over the effect of the smooth broad-band power. We measure the location of the BAO feature by fitting the observed auto (band) power spectrum $\Clobi{(\ell)}$\footnote{Note that we switched the notation from $p_{\beta}$ to $C(\ell)$.} with the following fitting formula using a template power spectrum $\Clmoi{(\ell/\al)}$:
\begin{eqnarray}
\Clobi{(\ell)}=B_{z_i}(\ell)\Clmoi{(\ell/\al)}+A_{z_i}(\ell), \label{eq:fiteq}
\end{eqnarray}
where $\al$, $B_{z_i}(\ell)$, and $A_{z_i}(\ell)$ are fitting parameters. The functional form for $B_{z_i}$ and $A_{z_i}$ is discussed in \S~\ref{subsec:BAs}. The parameter $\al$ measures the angular location of the BAO relative to that of the fiducial cosmology. That is,
\begin{eqnarray}
&&\al = \lobs/\lfid= [\DArs]_{\rm obs}/[\DArs]_{\rm fid},\label{eq:ratioal}
\end{eqnarray} 
where $[\DArs]_{\rm fid}$ is the fiducial angular location of the BAO in the template and $[\DArs]_{\rm obs}$ is the measured angular location of the BAO. A value of $\al > 1$ suggests that the observed angular location of the BAO is smaller than that of the fiducial cosmology.
For each redshift bin, $z_i$, free parameters $B_{z_i}(\ell)$ and $A_{z_i}(\ell)$ account for the smooth modification of the power spectrum from the template due to nonlinear structure growth and any scale-dependent bias. Finally, we use power spectra for \LRGs, \LRGe, \LRGn, and \LRGt\ simultaneously and fit for a universal $\al$ while marginalizing over $B_{z_i}$ and $A_{z_i}$ independently at each redshift bin. 

The template $\Clmoi{(\ell/\al)}$ is constructed from the 2-dimensional projection of 3D power spectrum \citep{Fisher94,Pad07}. Including linear redshift distortions,
\begin{eqnarray}
&&\Clmoi{}(\ell)=\frac{2}{\pi} \int \mathrm{d}k k^2 P_m(k,z_i) \nonumber\\
&&\left( \int \mathrm{d}z   \frac{\mathrm{d}N_i}{\mathrm{d}z}  b(z) \frac{D(z)}{D(z=0)}  \left[ j_\ell\left({r(k,z)}\right) -\beta j_\ell^{''}\left({r(k,z)}\right) \right] \right)^2,\label{eq:Clmode}
\end{eqnarray}
where $r(k,z)=k(1+z)\Daf$, $dN_{z_i}/dz$ is the normalized, true redshift distribution of galaxies for the corresponding $i^{th}$ photometric redshift bin (Figure \ref{fig:dndz}), $j_\ell$ is the spherical Bessel function, $j_\ell^{''}$ is the second derivative of the spherical Bessel function with respect to $r(k,z)$, $b(z)$ is a fiducial galaxy bias, $\beta$ is the fiducial redshift distortion parameter\footnote{$\beta =\Om^{0.56}(z)/b(z)$}, and $D(z)$ is the linear growth rate. \scr{Due to the projection, redshift distortions significantly affect the broad-band shape of the power spectrum only for $\ell < 30$; see the dashed and dotted lines in Figure \ref{fig:Pk} \citep[also see e.g.,][]{Nock10}.} 

Knowing $dN_{z_i}/dz$ precisely is critical for constructing the correct BAO location in the template. Note that, thanks to the extensive spectroscopic training set from the BOSS CMASS galaxies ($112,778$ objects), we have an excellent determination of the redshift probability distribution for our photometric luminous galaxy samples. Once we calculate the template $\Clmoi{}(l)$ for the fiducial cosmology, we rescale the wavenumber with $\al$ to generate $\Clmoi{}(\ell/\al)$ that fits the observed power spectrum in Equation \ref{eq:fiteq} \scr{while marginalizing over the free parameters $B_{z_i}(\ell)$ and $A_{z_i}(\ell)$}. 

The term $\Pm$ is generated by degrading the BAO portion of the fiducial, linear power spectrum with a nonlinear damping parameter $\Signl = 7.527 [D(z)/D(0)]\hMpc$ to mimic the nonlinear evolution of the BAO due to the structure growth \citep{ESW07}:
\begin{eqnarray}\label{eq:Pmodel}
\Pm(k,z_i)&=&\left[ \Plin(k)-\Psm(k)\right] \exp{\left[ -k^2 \Signl(z)^2/2 \right]} \nonumber \\
&&+\Psm(k),
\end{eqnarray}
where $\Plin$ is the linear power spectrum at $z=0$ derived from CAMB \citep{CAMB} and $\Psm$ is the nowiggle form, i.e., the power without BAO, calculated using the equations in \citet{EH98}. The smoothing of the BAO is dominated by the width of the underlying redshift distribution, and the exact choice of $\Signl(z)$ does not have a significant impact. 
For clarity we refer to $\Clmod$, instead of $\Pm$, as a `template' (angular) power spectrum and $\Pm$ as a `base' power spectrum in this paper.  We use a fiducial cosmology similar to the WMAP7 results \citep{Komatsu11} to define $\Pm$, $\Daf$, and etc: $\Om=0.274$, $\OL=0.726$, $h=0.7$, $\Ob=0.046$, $n_s=0.95$, and $\sig_8=0.8$. These values produce the fiducial sound horizon scale, $r_s=153.14\Mpc$, based on \citet{EH98}, and the fiducial angular location of the BAO, $[\DArs]_{\rm fid}=8.585$ at $z=0.54$.\footnote{\scr{Using the exact calculation of drag epoch rather than the fitting formula in \citet{EH98}, we find $r_s=149.18\Mpc$ for our fiducial cosmology, therefore $[\DArs]_{\rm fid}=8.813$ at $z=0.54$.} }

Note that, in equation \ref{eq:Clmode}, we are assuming that rescaling the sound horizon and angular diameter distance is equivalent to rescaling $\ell$ (i.e., the second equality of Eq. \ref{eq:ratioal}), which we call `$\alpha$ model'. This will be a reasonable approximation when the thickness of the redshift distribution that is projected on the 2D celestial surface is much larger than the scale of the clustering, i.e. in the limit of the Limber approximation. Based on the mock test in \S~\ref{sec:testassume} and the robustness test in \S~\ref{sec:results}, the $\alpha$ model appears to be a good enough approximation for our choice of fitting range and parameterization. 

The measured band power $\Clobi{(\ell)}$ has a contribution from a range of wave numbers, which is described by a window function (Eq. \ref{eq:Win}). Due to the large and contiguous survey area, the window function is sharply peaked with little correlation with neighboring bands (Figure \fc{windowf}). We account for this window function effect in the fitting. That is, for each wavenumber band $l_A$ for the redshift bin $z_i$, 
\begin{eqnarray}
&&\Clobi{(\ell_A)}= \nonumber \\
&& \times \int^{600}_1 [B_{z_i}(\ell')\Clmoi{(\ell'/\al)}+A_{z_i}(\ell')]W_i(\ell'|\ell_A)d\ell'. \label{eq:fiteqw}
\end{eqnarray}

\begin{figure}[b]
\plotone{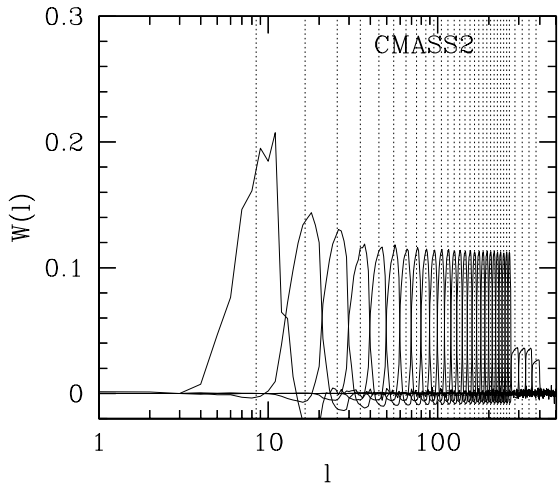}
\caption{Window function of \LRGe, as an example.}\label{fig:windowf}
\end{figure}

\subsection{A choice of $B(\ell)$ and $A(\ell)$}\label{subsec:BAs}
There are a few considerations to make when choosing the optimal parametrization for $B(\ell)$ and $A(\ell)$.  We want $B(\ell)$ and $A(\ell)$ to be flexible enough to model and remove the broad-band shape of the power spectrum. With the flexibility in $B(\ell)$ and $A(\ell)$, we also gain some tolerance on a possible, small difference between the BAO feature in the fiducial template and the observed feature by trading power between $B(\ell)\Clmod$ and $A(\ell)$. On the other hand, an arbitrarily flexible $B(\ell)$ and $A(\ell)$ will undesirably mimic BAO even with the no-BAO template. An extensive test of the parametrization for $B(\ell)$ and $A(\ell)$ for the template fitting method is discussed in \cite{SSEW08}, where they allow a large number of free parameters for $B$ and $A$ based on the spherically-averaged power spectra from \Nb\ realizations which correspond to a spectroscopic survey. With our photometric redshift uncertainty, a projection of the BAO from different distances introduces an additional damping in the feature, leaving higher harmonics other than the first much less distinct relative to the noise level. Therefore we are forced to limit the flexibility in $B(\ell)$ and $A(\ell)$ more strictly than the case assumed in \cite{SSEW08} while still making sure that the BAO scale is correctly recovered. We choose a revised fitting range so that the broadband is well modeled despite the smaller number of $B(\ell)$ and $A(\ell)$.

Based on tests with mock catalogs (\S~\ref{sec:testassume}), we use a fitting range $30<l<300$ and a linear function in $\ell$ for $B_{z_i}$ and a constant $A_{z_i}$ (abbreviated with `A0B1' hereafter):
\begin{eqnarray}
B_{z_i}(\ell)&=&B_{z_i0}+B_{z_i1}\ell \\ 
A_{z_i}(\ell)&=&A_{z_i0}.
\end{eqnarray}
Therefore, for four redshift bins, we fit for a total of 13 parameters including $\al$.

\subsection{Parametrizing $\DA$}
The observed location of the BAO peaks in a power spectrum is determined by the angular diameter distance at each redshift. In a sample selected using photometric redshift, the location of the BAO feature for a given redshift slice depends on an integration over the broad range of the true redshift distribution, rather than a distance at a single redshift, as evident in equation \ref{eq:Clmode}. In an ideal case, we may attempt to constrain the redshift dependence of $\DA$ over the entire spectroscopic redshift range in a non-parametric way \cite[e.g.,][]{Percival10} in the fitting process. In reality, the Fisher matrix analysis predicts an error of $\sim 4\%$ on a single measurement of $\DA$ for all our four redshift bins combined. Also while the expected true redshift distribution has long tails ($0<z_s<1$), the distribution tends to peak sharply for each photometric bin so that most of the information for a given bin is well concentrated within $\pm \sigma_{z_{\rm ph}}$. Third, for a reasonable range of cosmology, the shape of $\DA$ does not evolve significantly between $z=0.45$ and $z=0.65$. We therefore expect little gain for designing our analysis to measure multiple $\DA$s, i.e., the evolution of $\DA$ as a function of redshift. We instead design our fitting method to measure a single, more precise distance measurement at the redshift that contributes the most information. Based on the median and mean redshift of the photometric sample, we assign our measurement of $D_A$ to $z=0.54$.

In detail, we assume $\Daf$, given by the fiducial cosmology, and set 
\begin{eqnarray}
\DA = \al \Daf.\label{eq:daeq}
\end{eqnarray}
That is, we fix the shape of the $\DA$ to be the same as $\Daf$ and measure the amplitude of $\DA$. 

\subsection{Clustering evolution of lumious galaxies}
In generating the template $\Clmod$, we need to make a prior assumption on the evolution of the galaxy bias of LGs and the linear growth rate (Eq. \ref{eq:Clmode}). We consider two extreme cases of the galaxy clustering evolution: first, we assume that the overall clustering, $b^2D^2$, does not change with redshift, which we call as `{\it con-cluster}'. Second, we assume that the bias does not change with redshift, which we call as `{\it con-bias}'.  The two cases make little difference in the final best fit of $\al$, mainly because the expected true redshift distribution sharply peaks within $\pm \sigma_{z_{\rm ph}}$, compared to the galaxy clustering evolution. Note that, by marginalizing over $B_{z_i}$ at each photometric redshift bin $z_{i}$, we take into account the evolution of galaxy clustering across different redshift bins whether we use `con-cluster' and `con-bias'. As a default, we fix $b=2$ inside $\Clmod$ (i.e., `{\it con-bias}', and therefore the best fit $B_{z_i}$ can be {\it approximately} interpreted as $b^2(z_i)$.

\section{Testing the method}\label{sec:testassume}
Before applying our fitting method to the real data, we want to validate, using mock catalogs, that our fitting method returns an accurate estimate of the BAO scale. In other words, we want to check that neither our process of deriving optimal quadratic estimators of band powers nor our fitting method biases the measured BAO scale. 

\subsection{\Nb\ mocks}\label{subsec:Mocks}
\begin{figure}[t]
\centerline{\epsfxsize=3in\epsffile{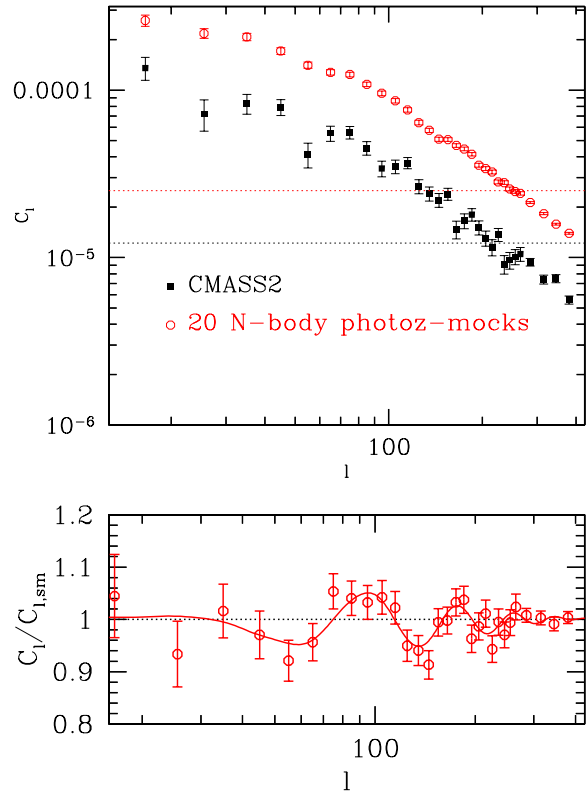}}
\caption{The red circles with error bars show a power spectrum averaged over 20 \Nb\ mocks for the same line of sight. The black points show the power spectrum of \LRGe. The dotted lines in the top panel show the shot noise contribution in both cases. The solid line in the bottom panel show the expected BAO feature.}\label{fig:MockPk}
\end{figure}

As explained in Paper I in detail, we generate mock catalogs of our imaging data making use of the 20 CMASS mocks constructed by \cite{White11}. We call these `\Nb\ photoz-mocks'. The cosmology used for generating these mocks
 is the same as our fiducial cosmology. The comoving volume of the original CMASS mock is $[1.5\hGpc]^3$  
and, to build \Nb\ photoz-mocks, we extract an octant of a spherical shell between $r=1.33\hGpc (z=0.5)$ and 
$1.45\hGpc (z=0.55)$ from the origin (one corner of a simulation box) and project galaxies along the radial direction 
without introducing photometric redshift errors. \scr{For simplicity,} we do not include 
redshift distortions, which will be visible only on very 
large scales (Figure \ref{fig:Pk}), nor the effect of the mask in generating these mocks. The resulting power spectrum of the projected 
field has a BAO feature that is quite similar to that expected for \LRGe (i.e., $0.5<z<0.55$).
Each of the resulting \Nb\ photoz-mocks spans a 
$\pi/2\;{\rm rad}^2\;(=5157\;{\rm deg}^2)$ and contains $\sim 125,000$ galaxies. While the number of galaxies is smaller than \LRGe, the amplitude of power spectrum is boosted overall by almost the same factor due to the thinner redshift slice: the signal-to-noise ratio per mode of each mock is therefore similar to \LRGe. (see Figure \ref{fig:MockPk}). Therefore the photoz-mocks serve as reasonable mocks for the observed power spectrum.

\begin{figure}[t]
\centerline{\epsfxsize=3in\epsffile{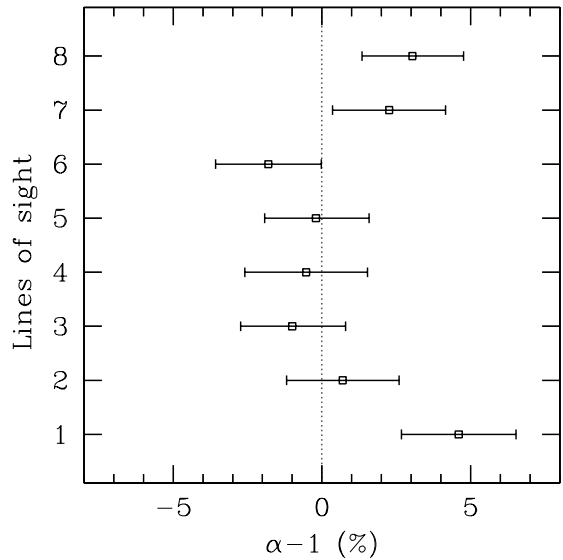}}
\caption{The data points show the best fit $\al-1$ for the \Nb\ photoz-mocks for different lines of sight. For each line of sight, the power spectrum is averaged over 20 \Nb\ photoz-mocks. Note that the different lines of sight are correlated.}\label{fig:Nbodyfit}
\end{figure}

\begin{figure*}
\centerline{\epsfxsize=6in\epsffile{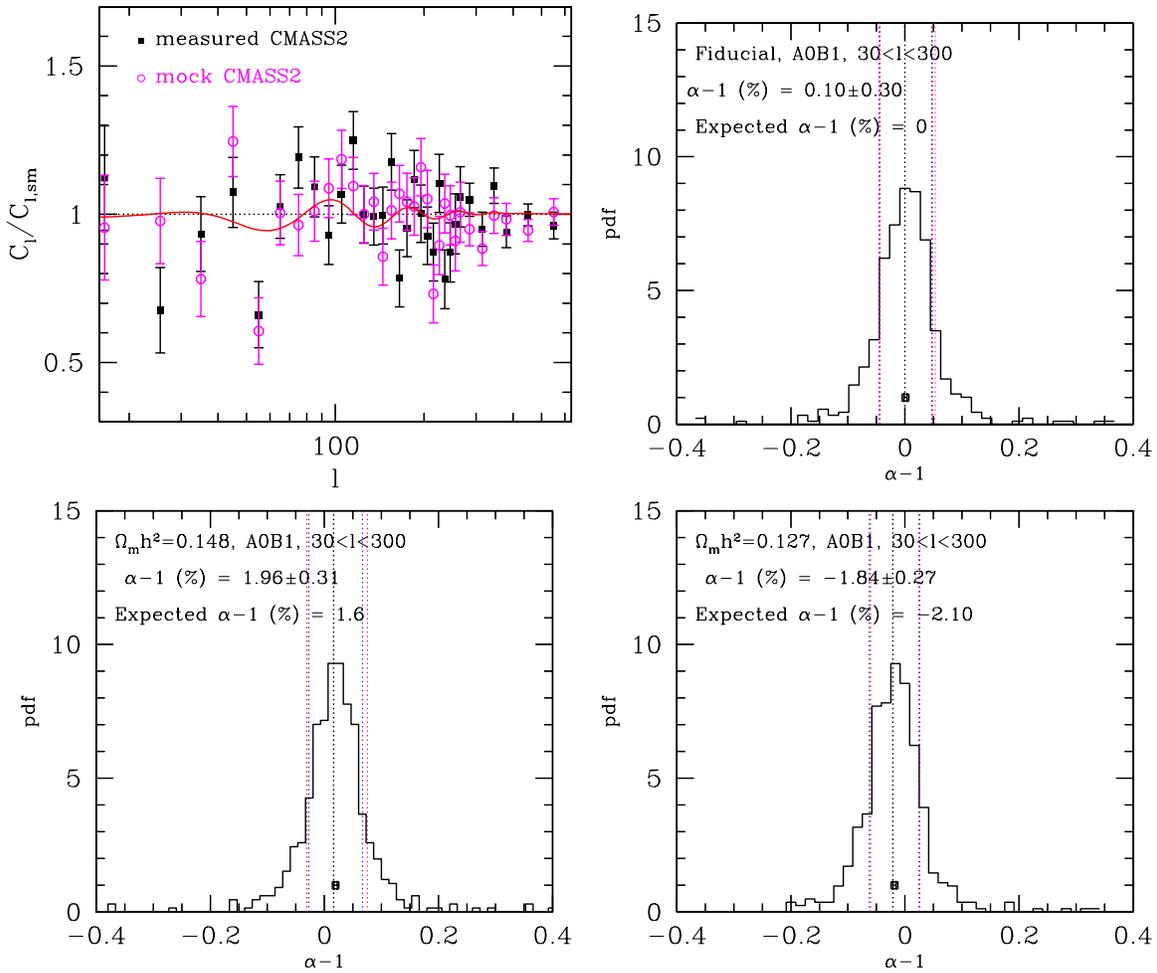}}
\caption{The top panel shows one of the Gaussian CMASS mocks for \LRGe\ (magenta circles) in comparison to the real data (black squares). The solid lines show the best fit model we used to generate the mocks.  The other three panels show the distribution of the 500 best fit $\al$ values of the Gaussian CMASS mocks. Top-right panel: the template is generated using the same cosmology as one used for generating the mocks (i.e., $\Oh=0.134$). The histogram shows the result using A0B1 over $30<l<300$. The black point with error bar shows the mean $\al-1$ and the error associated with the mean. The black dot-dashed line near the center shows the expected best fit $\al-1$ for each cosmology. The red and magenta (only in the top second panel) dotted lines show the {\it average error} derived from the 500 mock values for the range for $\Delta \chi^2 = \pm 1$ and the range for the 68.3\% of the likelihood, respectively. The blue dotted line shows the range that contains 68.3\% of the distribution of the 500 best fit $\al$s. The three values are quite similar. Bottom-left: the distribution of the best fit $\al$ of CMASS mocks when the template is built using $\Oh=0.148$. Bottom-right: using the template built assuming $\Oh=0.127$.}\label{fig:Gfour}
\end{figure*}

We repeat the procedure by placing an origin at eight different corners of each simulation box and generate eight sets (i.e., eight lines of sight) of 20 imaging mocks, i.e., a total of 160 mocks. Note that the eight lines of sight from each simulation box share a portion of volume and therefore are not independent of each other. 
 We generate the template power spectrum based on the galaxy distribution of the mock catalogs using Equation~\ref{eq:Clmode}.

To better detect a possible bias on $\al$ when using our fitting method, we increase the signal-to-noise ratio by averaging many power spectra.
We average power spectra of the 20 mocks for each configuration (i.e., each line of sight) and fit for $\al$. 
For the eight configurations, using `A0B1' (i.e., with $B_0+B_1\ell$ and $A_0$), we show the derived best fits and the 68.3\% range of the likelihood in Figure~\ref{fig:Nbodyfit}.
The distribution of eight $\al$ values appears slightly wider than what one would expect from a Gaussian distribution. However, it is not appropriate to compare this result with a Gaussian case, as the number of sets (i.e., eight) is too small to estimate the underlying distribution and the eight configurations are not independent from each other but share some portion of their volumes. Overall, we do not see an obvious indication \scr{either} that the best fit $\al$ from our OQE estimator is biased \scr{or that our fiducial fitting method introduces a bias}. 

\subsection{Gaussian CMASS mocks}\label{subsec:Gmocks}
We next test the accuracy of our fitting method using sets of mocks that have exactly the same noise properties as the real data. We generate many Gaussian mock power spectra using the covariance matrix for the real data described in \S~\ref{subsec:cov}. In detail, we find the best fit band powers for the power spectra measurements of \LRGs, \LRGe, \LRGn, and \LRGt\ over the entire wavenumber range (i.e., $1<l<600$), while deliberately using a slightly different choice of $B(l)$ and $A(l)$ (A2B0) than our fiducial method; $\al$ is fixed to be unity during this process. This approach allows us to construct theoretical power spectra for the fiducial cosmology while taking into account the realistic broad-band shape of LGs. Using the four best fit band power spectra and using the covariance matrix for the real data, we generate 500 sets of Gaussian CMASS mocks. For each set, the four CMASS mock power spectra therefore mimic \LRGs, \LRGe, \LRGn, and \LRGt\ not only in terms of clustering but also in terms of the covariance among them. One of the mock \LRGe\ is shown in the top left of Figure \ref{fig:Gfour} in comparison to the real data.

We then apply our fiducial fitting method to the 500 CMASS mocks. Since the mock data and the template use the same fiducial cosmology, we expect the average value of $\al$, if unbiased, to be unity. The top right panel of Figure \ref{fig:Gfour} shows the pdf distribution of the 500 best fit $\al$ values when we fit to the mock data over $30<l<300$ using A0B1. The mean and the standard deviation of $\al-1$ is $(0.10 \pm 6.60)\%$. After rescaling the standard deviation by the square root of the number of samples, the mean and the error associated with the mean value of $\al$ is $(0.10\pm 0.30)\%$, i.e., unbiased within $1-\sigma$. Decreasing the number of free parameters to A0B0 causes a biased estimation of $\al$ as large as $2\%$ in $\al$. \scr{We find A1B0 also gives an unbiased result, while we find that parametrizations with more free parameters result in a non-negligible degree of bias on the BAO location.}

The distribution has significant `tails' as shown in Figure \ref{fig:Gfour}. As a result, the standard derivation (6.60\%) is much larger than the range that contains 68.3\% of the distribution: $\al-1(\%)=\afit{0.1}{4.8}{4.3}$. That is, the distribution of $\al$ appears wider than the Gaussian one. We have excluded a few catastrophic outliers ($\sim 0.8\%$ of the samples for the fiducial case) that show $\al-1(\%) > 40$ in deriving these statistics. Obviously, these cases are not fitting to the BAO feature. Some of these catastrophic outliers show a factor of a few larger errors on $\al$ than the rest of the samples, while others show a reasonable error associated with the best fit. We find that the reduced $\chi^2$ of these catastrophic outliers are not necessarily large. While we exclude these extreme outliers so that the statistics of the distribution are not dominated by these occasions, we note that, in analyzing a real data, we expect to derive a very wrong value of BAO scale more likely than would be expected were this distribution perfectly Gaussian.

For the real data, we have only one realization and therefore need to check that the error we quote for the data will be closely approximating the $68.3\%$ range of the sample distribution if we had more than one sample. For each CMASS mock set, we derive errors associated with the best fit $\al$ by applying $\Delta \chi^2 = \pm 1$ or by deriving the  width that contains 68.3\% of the likelihood. On average, the resulting error on $\al$ for an individual CMASS mock is $\al-1(\%)=\afit{0.1}{4.7}{4.4}$ and $\afit{0.1}{5.2}{4.7}$ for $\Delta \chi^2$ of $\pm 1$ (dotted red lines in Figure \ref{fig:Gfour}) and the 68.3\% width of the likelihood (dotted magenta lines), respectively; the latter is slightly larger than the former. These are reasonably similar to the $68.3\%$ range of the best fit distribution of the mocks, which was $\al-1(\%)=\afit{0.1}{4.8}{4.3}$ (dotted blue lines). We therefore will quote the 68.3\% range of the likelihood surface as our formal error for the real data.

\subsection{Variations in the template}\label{subsec:vartem}
In Section~\ref{subsec:Gmocks} we assumed that the fiducial cosmology used for constructing the template matched the true cosmology, and we could recover an unbiased result in this situation. In fact we also need to confirm that our method is unbiased when the cosmology used in the analysis does not match the true cosmology.
The shape of the BAO feature is determined by the matter density and the baryon density, which is well measured by the current CMB observations. \cite{SSEW08} investigated how the results of the template fitting depends on small deviations in $\Pm$ and found that the effect is negligible (less than 0.02\% bias on $\al$). As explained in \S~\ref{subsec:BAs}, our method in this paper uses a smaller set of free parameters due to the lower signal-to-noise level of the real data \footnote{\cite{SSEW08} uses an \Nb\ volume of $320\trihGpc$, as a comparison.} and, as a result, it is possible that the fitting result is more sensitive to the deviation in $\Pm$. We therefore revisit this issue. Since $\Obhh$ is better measured than $\Oh$ by the CMB data, we only vary $\Oh$ from the fiducial value, while the current CMB constraint on $\Oh$ is $0.1326\pm0.0063$ \citep{Komatsu11}. In order to leave the shape of the angular diameter distance to redshift relation unchanged, we hold $\Om$ fixed and vary $h$ accordingly. 

The bottom left panel of Figure \ref{fig:Gfour} shows the distribution of the best fit $\al$ of 500 Gaussian CMASS mocks when the template is built using $\Oh=0.148$ (i.e., $10\%$ away from $\Oh=0.134$ used for the mocks). We again use the fiducial, A0B1 parameter set and a range of $30<l<300$. We find that using a smaller number of parameters, i.e., A0B0, makes the result more vulnerable to the variations in the template. Based on the sound horizon scale and $h$ in this cosmology, we expect $\al-1 (\%) = 1.6$ (the dashed vertical line). The average of the best fit is $\al-1 (\%) = 1.96 \pm 0.31$, and therefore we recover the correct BAO scale. The bottom right panel of Figure \ref{fig:Gfour} shows the distribution of $\al$ using $\Oh=0.127$, i.e., $-5\%$ away from the true value used for generating the mocks. We expect $\al-1 (\%)= -2.10$ while we measure $-1.84\pm 0.27$; we recover the expected value. We also have tested A1B0: we find a moderate bias for $\Oh=0.148$ due to asymmetric tails, while the bias is overall less than 0.6\%. This parametrization therefore would be a reasonable choice as well given the level of signal to noise of our data.
 
In the next section, we will apply the same test to the real data and show that the measured BAO scale does not change as a function of the fiducial cosmology assumed for the template. In addition to the assumption we made for $\Pm$, we have also assumed a fiducial relation of angular diameter distance to redshift. We will show that we recover the same result over a range of angular diameter distance to redshift relationship.

Summarizing our mock tests, we find that the angular power spectra produced by our OQE code \citep{Ho11} show no strong sign of a bias in its BAO feature and that `A0B1' is a good choice of parametrization over $30<l<300$ for our data quality and therefore returns the correct BAO scale within $0.3\%$ over a reasonable range of variations in our assumption.

\section{Results: DR8 imaging data}\label{sec:results}

\subsection{Building the covariance matrix}\label{subsec:cov}
We use Gaussian covariance matrices calculated for the auto power spectra of the four DR8 redshift bins using equation \ref{eq:Covab}. As shown in Paper I, we do not find an obvious indication that the Gaussian covariance matrix for the OQE estimator underestimates the true error of the 2-dimensional projection of the nonlinear galaxy field. Also \citet{Takahashi11} and \citet{Ngan11} have shown a negligible effect of non-Gaussian errors on the BAO measurement in multi-parameter fitting. We therefore retain the Gaussian field assumption when deriving the covariance between different redshift bins as well: we use the cross-power spectra between different redshift bins, while taking the window function due to the survey mask into account (see Paper I for more details). 

\begin{figure}
\centerline{\epsfxsize=2.41in\epsffile{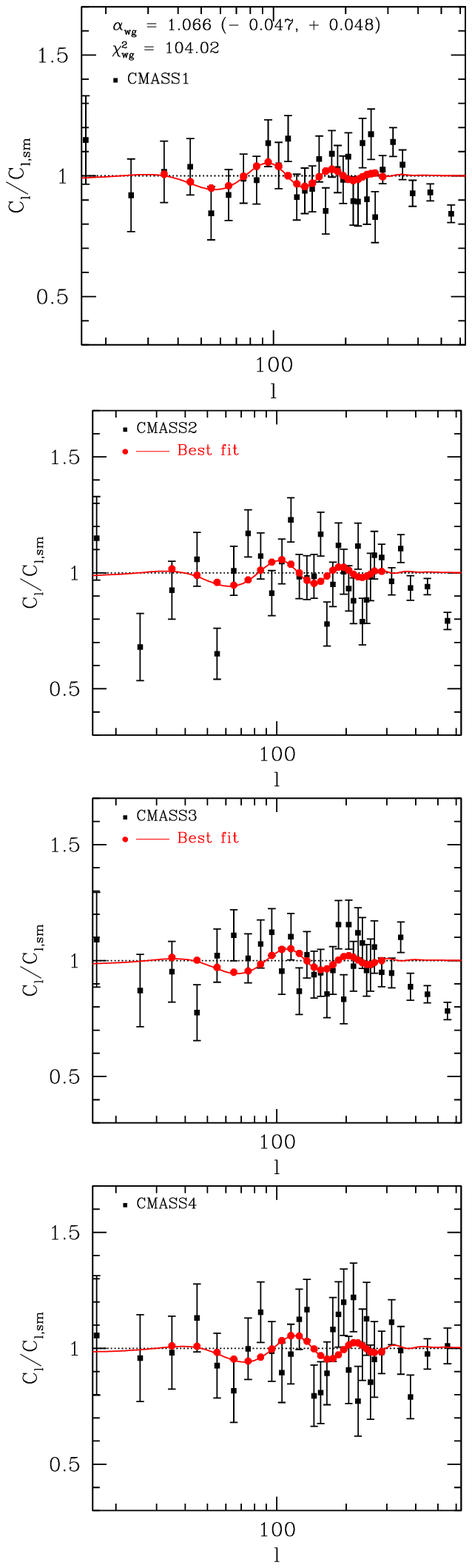}}
\caption{The best fit result using the combinations of \LRGs, \LRGe, \LRGn, and \LRGt\ and using A0B1. We derive $\al-1= \afit{6.609}{4.82}{4.68}\%$: i.e., 4\% deviation from the fiducial value based on WMAP7. The black data points in the four panels with error bars show the measured $C_l$ divided by a smooth fit at \LRGs, \LRGe, \LRGn, and \LRGt. The red lines show the resulting best fit $C_l$ and the red circles show the best fit band power $C_l$ after the window function effect is considered. }\label{fig:flrgall}
\end{figure}

\subsection{Best fit angular location of BAO}

\begin{figure*}
\centerline{\epsfxsize=6in\epsffile{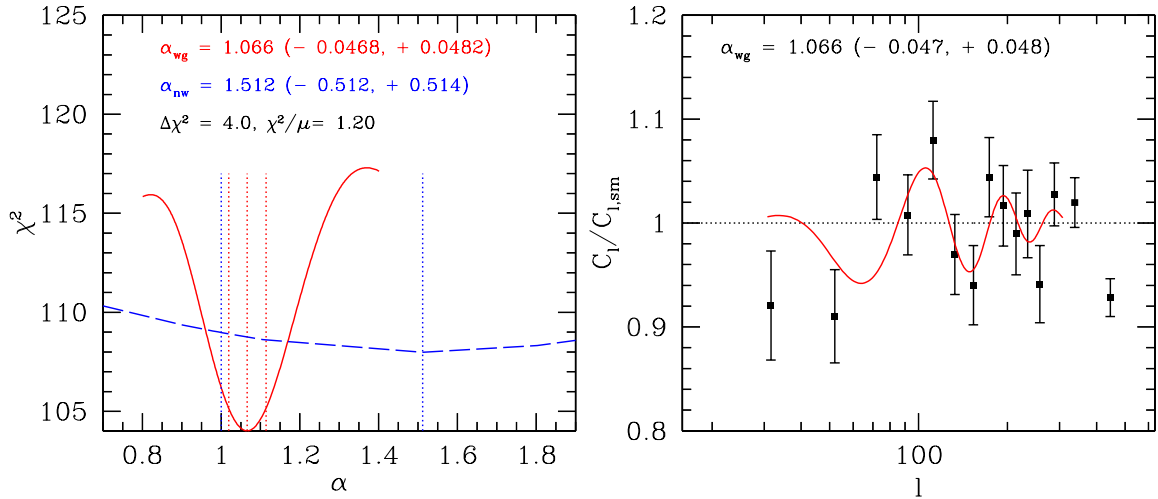}}
\caption{Left: the $\chi^2$ surfaces along $\al$ for Figure \fc{flrgall} when marginalized over other parameters (red line). The vertical dotted lines show the best fit $\alpha$ and the $1-\sigma$ range. \scr{The blue line shows the $\chi^2$ surface when the BAO is removed from the template.} \scb{Right: a stacked $C_l/C_{\rm l,sm}$ of the four panels of Figure \ref{fig:flrgall}. To better visualize the BAO feature we measured, we shift the wavenumbers of the four power spectra by $D_A(z_{\rm median})/D_A(z=0.54)$, re-bin the combined band powers while inversely weighting by errors. The solid red line is the best fit for \LRGe, after its wavenumber is rescaled to $z=0.54$.}  }\label{fig:flrgallchi}
\end{figure*}

We apply our fitting method to the DR8 imaging data and constrain the angular location of the BAO.
Figure \ref{fig:flrgall} shows our best fit result using the combinations of \LRGs, \LRGe, \LRGn, and \LRGt. The red lines/points show the best fit  $C_l$ with the BAO template in comparison to the measured data (black squares with error bars). The range of red points show the range of the fitting, i.e., $30<l<300$. In the figure, we denote the best universal fit $\al$ with the associated errors that correspond to the $68.3\%$ range of the likelihood distribution: we derive $\al-1 (\%) = \afit{6.61}{4.68}{4.82}$. The reduced $\chi^2$ at the best fit is $1.20$ for 87 degrees of freedom, and the probability of having a reduced $\chi$ value that exceeds this value is 10\%. The left panel of Figure \ref{fig:flrgallchi} shows the resulting $\chi^2$ surface along $\al$ when marginalized over other parameters (red line). Note that, due to the oscillatory feature of the BAO both in the data and the template, there are local minima around the global minimum of $\chi^2$. As implied in the figure by the extent of the red line, when we derive the $68.3\%$ range of the likelihood, we only include $\chi^2$ over $0.08<\al<1.4$, avoiding the local minima beyond this range. The right panel of Figure \ref{fig:flrgallchi} shows a stacked $C_l/C_{\rm l,sm}$ of the four panels of Figure \ref{fig:flrgall}. We perform this stacking procedure as follows. To better visualize the BAO feature we measured, we shift the wavenumbers of the four power spectra by $D_A(z_{\rm median})/D_A(z=0.54)$, where $D_A(z_{\rm median})$ is the median redshift for each redshift bin. We combine the four band powers, re-bin the combined data while inversely weighting each band power by its error. The solid red line is the best fit for \LRGe\ after its wavenumber is rescaled to mimic a result at $z=0.54$.

\begin{figure}
\plotone{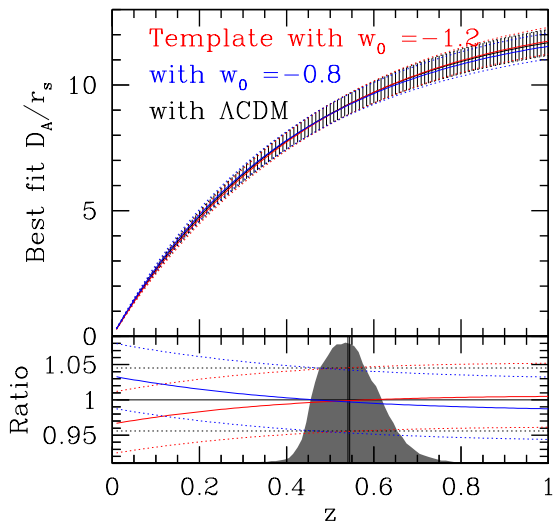}
\caption{Interpreting $\al-1(\%)= {6.61}$. Strictly speaking, the best fit value of $\al$ represents a constant ratio of the observed $\DArs$ to the fiducial $\DArsf$ we assume in the template. The black solid and dashed lines (with a shade) show what the best fit and the $1-\sigma$ error on $\al$ imply in this strict interpretation. However, although the redshift dependence of $\DA$ we assume spans $z\sim 0-1$, most of the galaxies are within $z\sim 0.45$ and $0.65$ with a peak of the distribution near $z \sim ~0.5-0.55$ (the shaded region in the bottom panel). Therefore, it is reasonable to consider that the best fit $\al \DArsf$ represents $\DArs$ near $z \sim ~0.5-0.55$. We also show the best fit using different template cosmologies: using $w = -0.8$ and $-1$. The bottom panel shows the ratios of different $\DArs$: $D_A/r_s/\DArsf$. One sees that the three different templates have a very similar shape in $\DA(z)$ over $z=0.45-0.65$, once the absolute difference is absorbed into $\al$. The three templates return virtually the same $\DArs$ at $z=0.54$.}\label{fig:variousda}
\end{figure}

Interpreting $\al-1(\%)= {6.61}$ requires our determination of the redshift to which this measurement corresponds. Strictly speaking, the best fit value of $\al$ represents a constant ratio of the observed $\DArs$ to the fiducial $\DArsf$ assumed in the template. The black solid and dashed lines (with a shade) in Figure \ref{fig:variousda} show what the best fit and the $1-\sigma$ error on $\al$ imply in this strict interpretation. However, although the redshift dependence of $\DA$ we assume spans $z\sim 0-1$, most of the galaxies are within $z= 0.45$ and $0.65$ with a peak of the distribution near $0.5<z<0.55$. Therefore, it is reasonable to consider that the best fit $\al \DArsf$ represents $\DArs$ near $z = 0.5-0.55$.
 The median and the mean of the weighted galaxy distribution are 0.541 and 0.544, respectively. We therefore adopt $z=0.54$ as the characteristic redshift that our BAO measured scale represents.

 To show that the best fit BAO scale indeed does not depend on the cosmology we assume for $\DArsf$, we repeat our fitting with templates constructed using different cosmologies. In detail, we vary the equation of state of dark energy, $w$, by $\pm 0.2$, such that $\DArsf$ at $z=0.54$ varies by $\sim 3.4-3.7\%$, i.e., slightly less than the $1-\sigma$ range associated with the best fit $\al$. In other words, we are testing the consistency of our answer by varying the template by $\sim 1-\sigma$ from the fiducial case of $\al$. Using a template with $w=-1.2$ , the best fit gives $\al-1= \afit{2.98}{4.81}{4.48}\%$ and, with $w=-0.8$, $\al-1= \afit{10.21}{5.02}{4.85}\%$. 

The top panel of Figure \ref{fig:variousda} shows the best fit $\DA/r_s$ ($=\al\DA/r_{s,\rm fid}$) using the three different templates. The solid lines show the best fit and the dotted lines show a  $1\sigma$ range of $\DA/r_s$. The bottom panel displays ratios of the best fit $\DA/r_s$ using no-$\LCDM$ templates with respect to the best fit $\DA/r_s$ using our fiducial $\LCDM$ template. From the top and bottom panel, one sees that the three different templates have a very similar shape in $\DA$ over $z=0.45-0.65$, once the absolute difference is absorbed into $\al$. The three templates return virtually the same $\DArs$ at $z=0.54$: they are consistent within $0.3\%$. We therefore quote the best fit using our fiducial $\LCDM$ template as our official measurement: $\DArs= \afit{9.212}{0.416}{0.404}$ at $z=0.54$. Using the current WMAP7 constraint on the sound horizon at drag epoch, $153.2 \pm 1.7 \Mpc$, we derive angular diameter distance $\DA= 1411\pm 65 \Mpc$ at $z=0.54$ \footnote{\scr{Using the exact integration rather than the fitting formula in \citet{EH98}, $r_s=149.18\Mpc$ and $\DA=\afit{9.456}{0.427}{0.415}$ at $z=0.54$.}}. Table \ref{tab:fitda} summarizes our best fit BAO location and the derived distance scale.

\begin{deluxetable*}{c|c|c}
\tablewidth{0pt}
\tabletypesize{\small}
\tablecaption{\label{tab:fitda}Best fit distance scale.}
\startdata \hline\hline
z & 0.54 \\ \hline
Assumptions & $r_{s, \rm fid}$ & $153.14\Mpc$ \\
& &  using \citet{EH98} \\ 
& $\DArsf$ & 8.584  \\\hline
Results& Best fit $\alpha$ & $\afit{1.0661}{0.0482}{0.0468}$  \\
& Best fit $\DArs$ & $\afit{9.212}{0.416}{0.404}$  \\ 
& Best fit $\DA$ & $1411 \pm 65 \Mpc$ \\
&& using the prior, $r_s$ = $153.7 \pm 1.7 \Mpc$ (WMAP7)
\enddata
\tablecomments{Our best fit BAO scale and the derived distance scale.}
\end{deluxetable*}

\begin{figure}
\plotone{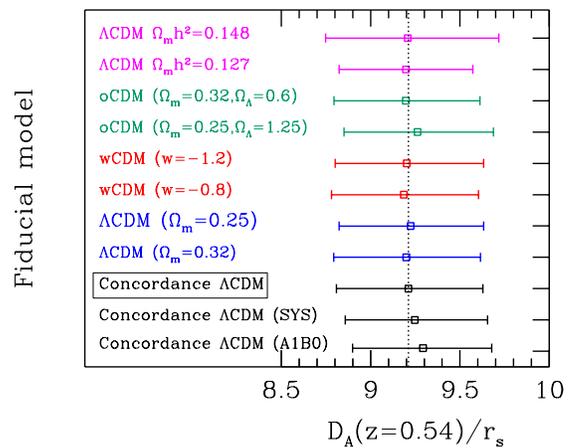}
\caption{The best fit $D_A(z=0.54)/r_s$ for various template cosmologies. We also show the result using a different parametrization (labeled with `Concordance $\LCDM$ (A1B0)' using A1B0) and the result with systematics correction (labeled with `Concordance $\LCDM$ (SYS)'). The one in the square box and the dotted vertical line are our fiducial choice: $\DArs= \afit{9.212}{0.444}{0.431}$ at $z=0.54$. The results are consistent within $1\%$.}\label{fig:robustda}
\end{figure}

We further test the robustness of our result by constructing templates using various cosmologies. Figure \ref{fig:robustda} shows the best fit $\DArs$ at $z=0.54$ assuming $w$CDM, $o$CDM, $\LCDM$, and assuming different values of $\Oh$. For the range of cosmologies we have investigated in this paper, the best fit varies less than $1\%$ in the acoustic scale while the $1\sigma$ error is $\sim 4.7\%$. The errors vary slightly more than the variations in the best fit, especially when the template cosmology deviates substantially from the concordance cosmology. \scr{We also test a different parametrization than the fiducial choice, A1B0, which has marginally passed the mock test (i.e., a likely bias of $\sim + 0.6\%$ on $\al$ based on the result in \S~\ref{subsec:vartem}). A1B0 gives  $\al-1=  7.5(\%)$, which is consistent with the fiducial result within $1\%$.}

If we remove the BAO in the template (the blue line in Figure \ref{fig:flrgallchi}), we essentially fail to constrain $\al$. This means that the flexibility in our fitting is sufficient that the broadband shape information cannot constrain $\al$. Therefore, we conclude that our measurement of $\DArs$ is mainly from the BAO information.

The precision of our measurement is much better than that of \citet{Carnero11}, where they measure the BAO location within 9.7\%. The discrepancy arises partly because of the larger survey-area coverage in this work, but the main difference is due to the difference in the modeling. \citet{Carnero11} assume a quite general model without utilizing the true redshift distribution of their photometric galaxies, and account for the various scenarios of the deviation between their model and the real data in their error budget. Our method is, on the other hand, a template-based approach utilizing the true redshift distribution of our photometric galaxies that is quite well determined by the extensive training set and generate a template with a precise BAO location given the redshift distribution and cosmology. Our method therefore is quite immune to the most of the systematic errors they account for. We observe approximately 1\% of variations depending on the choice of fiducial cosmology and parametrization, which has very small effect when added quadratically to our current error of 4.7\%.

\subsubsection{Significance of detection}
A reasonable concern regarding our measurement is whether or not we have fitted to only the BAO feature, or whether the result is offset due to noise spikes, which is obviously related to the significance of detection. The conventional method of determining significance of BAO detection is to use a template with (i.e., BAO template) and without BAO information (i.e., no-BAO template) and observe the difference between $\chi^2$ values of the two best fits. Unfortunately, such $\Delta \chi^2$ is very model-dependent. Using A0B1, we derive $\Delta \chi^2=4$, which can be conventionally interpreted as a $2\sigma$ detection of BAO. However, such detection level depends on the choice of parametrization.

We reconsider this issue of the detection level. Various observations including WMAP7 \citep{Komatsu11} and galaxy surveys \citep[e.g.,][]{Blake11b}, have already shown that BAO feature exists.
Given the signal-to-noise ratio level of our data, we are interested in how likely we have fitted to a BAO feature not a noise feature, rather than detecting the existence of BAO. Therefore, rather than fitting the power spectra with a no-BAO template, we shall fit \scr{many realizations of} the power spectra with a BAO template and see how often we derive the correct BAO scale. For a large sample variance, the BAO feature in the power spectrum may be wiped out by noise \citep[e.g.,][]{Cabre11}. Our mock test, Figure \ref{fig:Gfour}, shows that, in the presence of the sample variance that is the same as our data, we recover the true acoustic scale within $\sim 4.6\%$ in 68.3\% of the time. Obviously, we are not fitting to a BAO feature in the tails of the distribution.
 We therefore rephrase our detection level: our measurement is likely to recover the true BAO scale within $4.6\%$ in 68.3\% of cases, assuming that BAO exists. 

\subsection{Effect of systematics}
A number of observational systematics can potentially contaminate the observed galaxy clustering: stellar contamination, seeing variations, sky brightness variations, extinction, and color offsets \citep{Schlafly10}. However, as long as the systematics do not introduce a preferred scale similar to the BAO scale, i.e., if the systematics only introduce a smooth component in the power spectrum \scr{up to a sample variance}, our results would not depend on the contamination from systematics. Paper I moreover has shown that the effect of the survey systematics are small. We therefore have not included the systematic corrections for our main result. In this section, however, we use power spectra that were corrected for the systematics using the method introduced in Paper I \citep[See ][for a similar method for the correlation function]{Ross11} and observe the effect of the systematics on the result.

The method in Paper I assumes that the effect of systematics is small and linear. In Fourier space, therefore, we assume that the following equation holds for each wave band $\ell$:
\begin{equation}
\hat{\delta}_{z_i}(\ell)  = \hat{\delta}_{g,z_i}(\ell)+\sum_{s_a} \epsilon_{z_i,a}(\ell)\hat{\delta}_{s_a}(\ell),
\end{equation}
where $\hat{\delta}_{z_i}(\ell)$ is the observed galaxy density field at the $z_i^{\rm th}$ redshift bin, $\hat{\delta}_{g,i}$ is the true galaxy density field, and $\delta_{s_a}$ are the variation of systematics across the sky. We only include the dominant three systematics identified in Paper I: stellar contamination, seeing variations, and sky brightness variations. If we assume that there is no intrinsic correlation between the systematics and the underlying large scale structure, i.e., $<\hat{\delta}_{g,z_i} \hat{\delta}_{s,a}(\ell)> = 0$, we can solve for $\epsilon_{z_i,a}$ using the measurements of galaxy power spectra (i.e., $<\hat{\delta}_{z_i}  \hat{\delta}_{z_j}>$) and the cross-power spectra between galaxies and the systematics (i.e., $<\hat{\delta}_{z_i}  \hat{\delta}_{s_a}>$), as presented in Paper I. The error on the band power is minimally propagated: the error is quadratically increased by the amount of the final correction, after taking into account the number of wave modes. 

\begin{figure}
\centerline{\epsfxsize=2.41in\epsffile{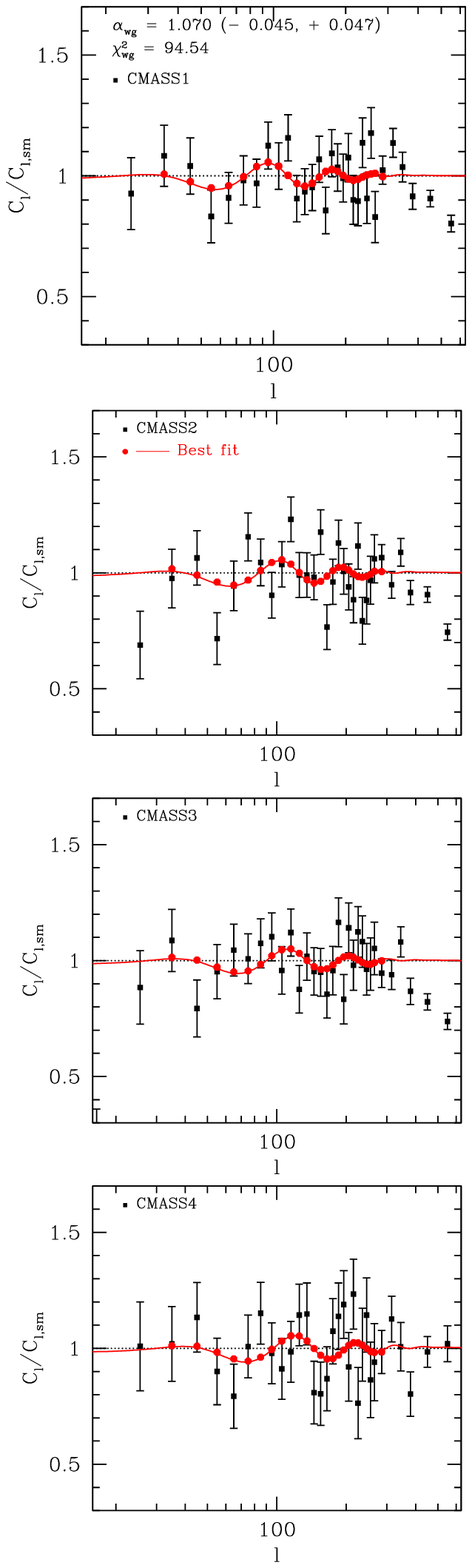}}
\caption{The best fit result after systematics correction using the combinations of \LRGs, \LRGe, \LRGn, and \LRGt\ and using A0B1. We derive $\al-1= \afit{7.012}{4.71}{4.51}\%$, which is quite similar to the result before systematics correction. The black data points in the four panels with error bars show the measured $C_l$ after systematics correction divided by a smooth fit at \LRGs, \LRGe, \LRGn, and \LRGt. The red lines represent the best fit $C_l$ and the red circles show the best fit band power $C_l$ after the window function effect is considered. }\label{fig:flrgallSys}
\end{figure}

\begin{figure*}
\centerline{\epsfxsize=6in\epsffile{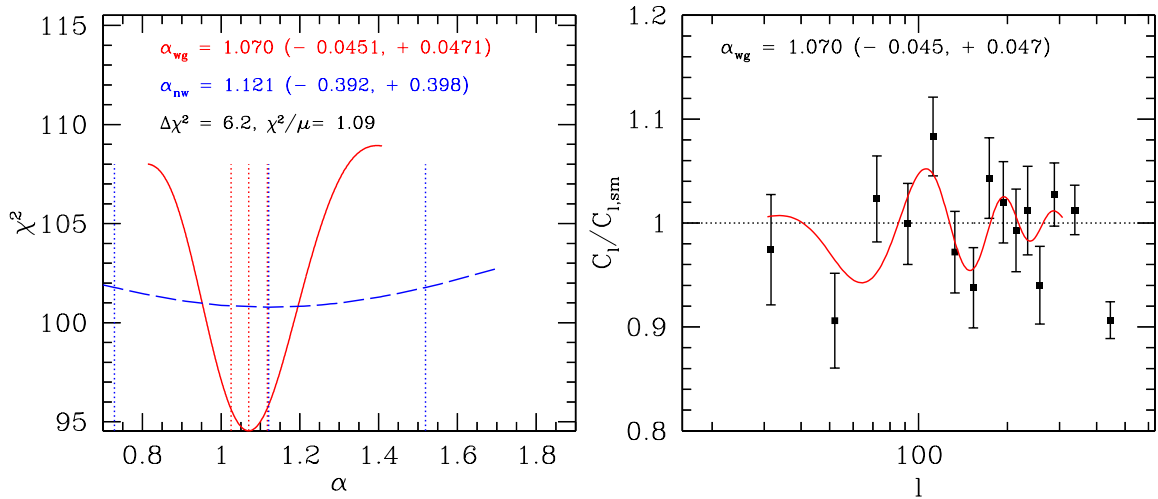}}
\caption{Left panel: The $\chi^2$ surfaces along $\al$ after systematics correction for Figure \fc{flrgallSys} when marginalized over other parameters (red line). \scr{The blue line shows the $\chi^2$ surface when the BAO is removed from the template.} \scb{Right panel: a stacked $C_l/C_{\rm l,sm}$ of the four panels of Figure \ref{fig:flrgallSys} after systematics correction. To better visualize the BAO feature we measured, we shift the wavenumbers of the four power spectra by $D_A(z_{\rm median})/D_A(z=0.54)$, re-bin the combined band powers while inversely weighting by errors. The solid red line is the best fit for \LRGe, after the wavenumber is rescaled to $z=0.54$.}}\label{fig:flrgallchiSys}
\end{figure*}

Figure \ref{fig:flrgallSys} and \ref{fig:flrgallchiSys} show the best fit results when we use the power spectra after systematics correction. We derive $\al-1= \afit{7.012}{4.71}{4.51}\%$. The reduced $\chi^2$ has slightly improved to be 1.09. The difference in $\chi^2$ between using the BAO template and the no-BAO template has increased to 6.2 after systematics correction, from the previous 4.0 without systematics correction. \scb{The right panel of Figure \ref{fig:flrgallchiSys}, in comparison to Figure \ref{fig:flrgallchi}, shows that the systematics correction, while the effect is small, improves the fit on large scales $(l < 100)$. }

Figure \ref{fig:robustda} shows that the best fit value with the systematics correction is consistent with the fit before the systematics correction within 1\% of $\al$, demonstrating that the BAO fitting is fairly robust against the systematics effects. The overall improvement in the statistics after the systematics correction, such as on the reduced $\chi^2$, motivates the usage of the method in Paper I for future surveys, which can be further improved with a more careful error propagation during the correction.

\section{Discussions: Cosmological implications}\label{sec:discussions}
\begin{figure}
\centerline{\epsfxsize=3in\epsffile{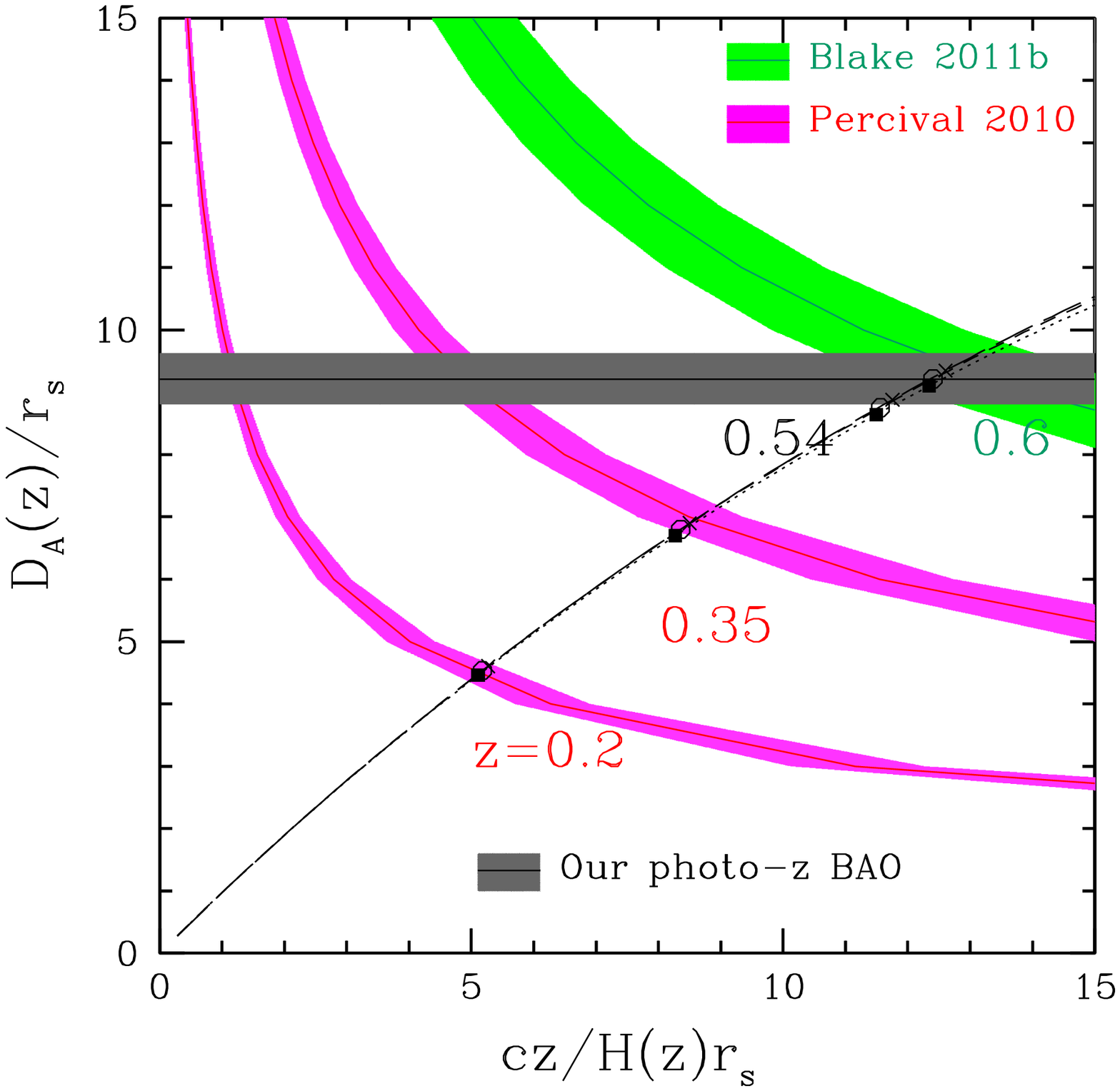}}
\caption{Various BAO measurements in comparison to the concordance $\LCDM$. The measurement of $D_A(z=0.54)$ from this paper is shown with the black horizontal line. The gray shade represents  $1-\sigma$ error. Red lines with magenta shades show $\DV(z=0.2)/r_s$ and $\DV(z=0.35)/r_s$ from \citet{Percival10}
 and the green line shows $\DV(z=0.6)/r_s$ from \citet{Blake11b}. The black squares along the diagonal dotted line show the expected combination of $\DA$ and $H$ based on the concordance $\LCDM$ at the redshifts of the data. One sees that the data beyond $z=0.35$ observed the BAO at a slightly smaller scale (i.e., a larger distance) than the concordance $\LCDM$. The circles along the dashed line and the crosses along the dot-dashed line show the expected $\DA$ and $H$ based on the best fit $w$CDM and $o$CDM cosmologies in Table \ref{tab:cosmomc}.}\label{fig:baoall}
\end{figure}

We combine our measurements of $D_A(z=0.54)$ with recent spectroscopic BAO measurements. The spectroscopic surveys report $\DV(z)$ that contains both the information along the line of sight, $H(z)$, and the information on the transverse direction, $D_A$. In Figure \ref{fig:baoall}, we present the distance-to-redshift relations of different BAO measurements in a 2-dimensional space of $D_A(z)$ and $H(z)$. Our measurement of $D_A(z=0.54)$ appears as the black horizontal line with the shaded region representing the associated error.  We also show the measurements of $\DV(z=0.2)/r_s$ and $\DV(z=0.35)/r_s$ from \citet{Percival10} for SDSS DR7 \citep{DR7} as red lines with magenta shades and $\DV(z=0.6)/r_s$ from \citet{Blake11b} for the WiggleZ data over $0.2 < z < 1$ as a green line with a light green shade. The black square points (along the dotted line) show the expected $\DA$ and $H$ at $z=0.2$, 0.35, 0.54, and 0.6 for our fiducial $\LCDM$. Note that the measurements beyond $z=0.35$ have a tendency to imply the location of the BAO at a smaller scale than the concordance $\LCDM$ (i.e., a larger $\DA$ than the fiducial cosmology), including our $D_A$ measurement ($\sim 1.4\sigma$ away). Due to nonlinear structure formation and galaxy bias, we expect about a $\sim 0.5\%$ of bias towards a smaller value on the measured BAO scale \citep{Crocce08,Pad09, Seo10, Mehta11}, which has not been accounted in these measurements. Such correction will slightly improve the consistency between the BAO measurements and the concordance $\LCDM$, but it is overall a very small effect for the current level of errors. The circles along the dashed line and the crosses along the dot-dashed line in Figure \ref{fig:baoall} show the expected $\DA$ and $H(z)$ based on our best fit $w$CDM and $o$CDM cosmologies from COSMOMC \citep{cosmomc02} that will be explained below.

\begin{deluxetable}{c|cc}
\tablewidth{0pt}
\tabletypesize{\footnotesize}
\tablecaption{\label{tab:cosmomc} The derived cosmological parameters.}
\startdata \hline\hline
 & $w$CDM & $o$CDM \\ \hline
$\Om$ & $0.2912$ $(0.2917) \pm 0.0270$ & $0.2939$ $(0.2952) \pm 0.0170$ \\
$h$  & $0.6884$ $(0.6892) \pm 0.0392$ & $0.6748$ $(0.6715) \pm 0.0175$ \\
$w$ & $-1.0185$ $(-1.0337) \pm 0.1862$ &   Fixed at $w=-1.0$\\
$\OL$ & $0.7088$ $(0.7083) \pm 0.2705$ & $0.7118$ $(0.7116) \pm 0.0172$ \\
$\Ok$ & Fixed at $\Ok=0.0$ & $-0.0057$ $(-0.0067) \pm 0.0058$  
\enddata 
\tablecomments{Marginalized fit and errors associated with the fit \scr{on selective parameters} that are derived using COSMOMC \citep{cosmomc02} for two different cosmologies. The value inside the parentheses show the best fit values. }
\end{deluxetable}

We use COSMOMC \citep{cosmomc02} to combine BAO measurements from the various galaxy surveys with the WMAP7 data \citep{Komatsu11} to derive constraints on cosmological parameters. For BAO measurements, we use $\DV(z=0.2)/r_s$ and $\DV(z=0.35)/r_s$ from SDSS DR7 \citep{Percival10},  $\DV(z=0.44)/r_s$, $\DV(z=0.60)/r_s$, and $\DV(z=0.73)/r_s$ from WiggleZ\footnote{For COSMOMC, we use the three-redshift slice representation of the WiggleZ data from \citet{Blake11b}, i.e., $0.2<z<0.6$, $0.4<z<0.8$, and $0.6<z<1.0$, accounting for the covariance among them, while in Figure \ref{fig:baoall} we show the result for the whole redshift range ($0.2<z<1.0$). \scr{Note that the distance measurements from the WiggleZ data include non-BAO information.}}, and $\DA(0.54)/r_s$ from this work. The WMAP7 data provides the sound horizon scale and the distance to the last scattering surface and therefore, in combination of the BAO measurements from the galaxy surveys, we can break the degeneracies and constrain $w$ and $\Omega_m$ (for $w$CDM) or $\Omega_\Lambda$ and $\Omega_m$ (for $o$CDM). \scr{The cosmological parameters that the COSMOMC chain vary are $\Obhh$, $\Omega_{c}h^2$, $\theta$, $\tau$, $n_s$, $\ln A_s$, and $A_{SZ}$, in addition to $w$ (for $w$CDM) or $\Ok$ (for $o$CDM); here, $\Omega_{c}h^2$ is the dark matter density, $\theta$ is the  approximate ratio of the sound horizon scale to the angular diameter distance to recombination, $\tau$ is the optical depth to reionization, $A_s$ is the primordial superhorizon power in the curvature perturbation on $0.05\Mpc^{-1}$ scales, and $A_{SZ}$ is the amplitude of the SZ power spectrum.} 

\begin{figure*}
\centerline{\epsfxsize=6in\epsffile{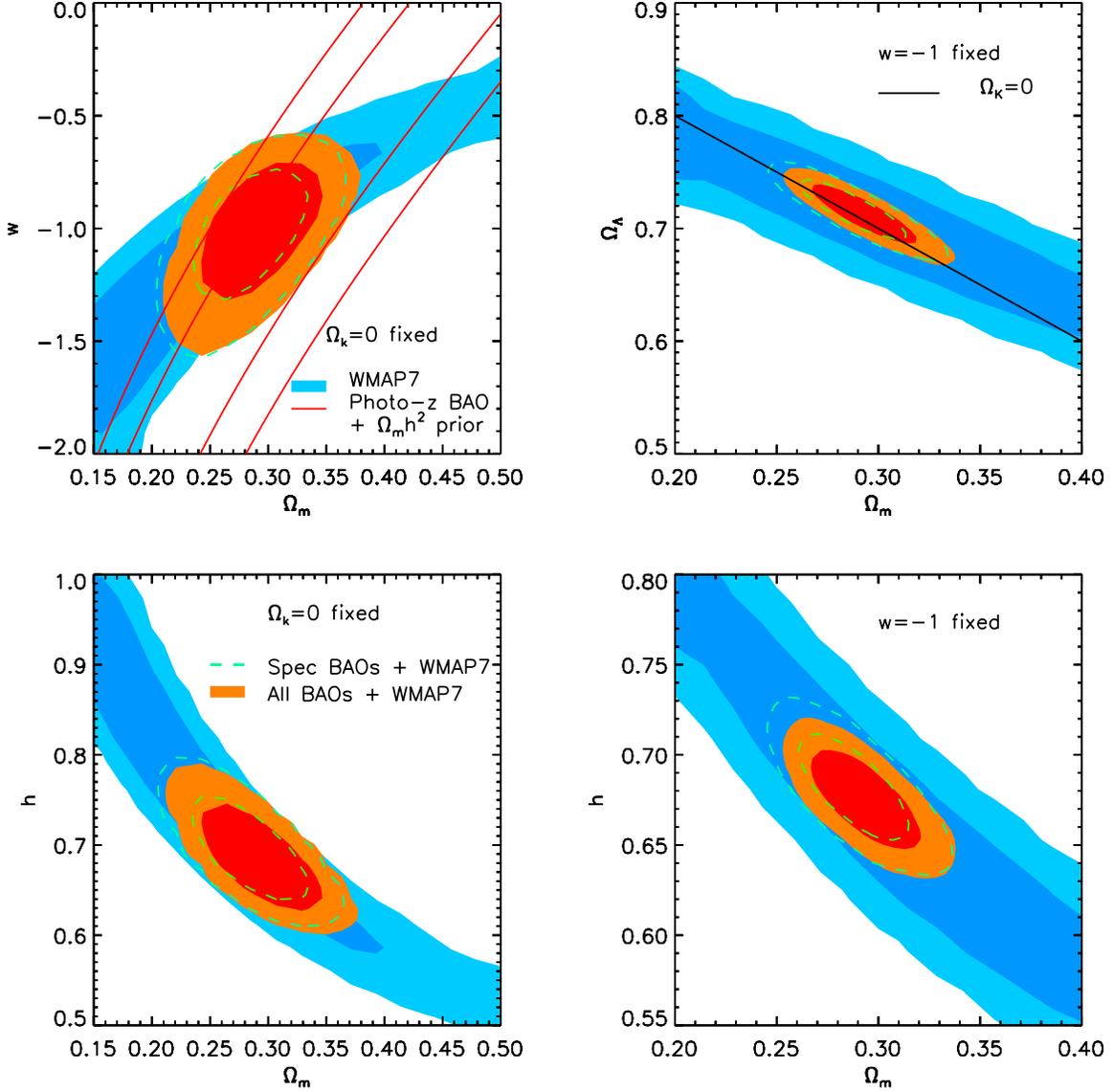}}
\caption{Reddish contours: constraints on cosmological parameters after combining all BAO measurements shown in Figure \ref{fig:baoall} and the WMAP7 constraints (blue shaded contours). The green lined contours show the constraints without our measurement. The left panel shows flat $w$CDM case and the right panel shows $o$CDM with $w=-1$. The reddish lined degenerate contours in the top left show the constraint from our data alone when using the current CMB prior on $\Oh$ (i.e, $0.1326\pm 0.0063$). The black line in the top right panel shows $\Ok=0$. }\label{fig:womh}
\end{figure*}

The left panels of Figure \ref{fig:womh} show marginalized 2-D likelihood contour surfaces that enclose 68.3\% and 95.5\% of the likelihood (reddish shaded contours) on $\Omega_m$ and $w$ (top) and $\Omega_m$ and $h$ (bottom) assuming a flat $w$CDM, in comparison to the case without our measurement (dashed green lined contours for the spectroscopic BAO measurements). The reddish contour lines in the top left show the constraint from our measurement alone using the current CMB prior on $\Oh$ (i.e, $0.1326\pm 0.0063$). This contour implies that, given the strong prior on $\Oh$, adding our measurement of the distance scale at $z=0.54$, which is larger than what is expected in the concordance $\LCDM$, weighs toward a slightly larger $\Om$ and therefore a slightly smaller $h$ with respect to the other data sets. We present the marginalized and the best fits of selective cosmological parameters in Table \ref{tab:cosmomc}: $\Omega_m = 0.2912 \pm 0.0292$, $w =-1.0185  \pm 0.186$, $h=0.6884 \pm 0.0392$ for a flat $w$CDM. The right panels show the 2-D contour on $\Omega_m$ and $\Omega_\Lambda$ for $o$CDM while holding $w=-1$. The best fit parameters are $\Omega_m=0.2939 \pm 0.0170 $, $\Omega_K=-0.0057\pm 0.0058$, and $h=0.6748\pm 0.0175$ in this case. Overall, an addition of our measurement slightly  increases $\Omega_m$ and decreases $\Omega_K$ toward a more negative value. In terms of errors, including our data point provides only a slight improvement on $\Om$ and $h$ for $o$CDM.

\section{Conclusion}\label{sec:con}
We have measured the acoustic scale from the SDSS-III DR8 imaging catalog using $872,921$ galaxies over $\sim 10,000 {\rm deg}^2$ between $0.45<z<0.65$. Galaxies are binned into four different redshift slices where the width of each slice is 0.05, which is approximately the error associated with photometric redshift determination. Angular power spectra are generated using an optimal quadratic estimator, as presented in Paper I. We use $\sim 110,000$ SDSS III BOSS galaxies as a training sample to derive the true redshift distribution of the galaxies in the imaging catalog and therefore build reasonable template power spectra. We fit the templates to the measured angular power spectra and derive the best fit acoustic scale while marginalizing over sufficient free parameters to exclude any non-BAO signal.  

We derive $\DArs= \afit{9.212}{0.404}{0.416}$ at $z=0.54$. Using the current WMAP7 constraint on the sound horizon at drag epoch, $153.2\pm1.7 \Mpc$, we derive angular diameter distance $\DA= 1411\pm 65 \Mpc$ at $z=0.54$. Without a BAO feature in the template power spectrum,  we cannot constrain a distance scale; the distance information we derive is therefore dominated by the BAO feature for our choice of parametrization.

Our measurement of the distance scale is quite insensitive to the fiducial cosmology we assume for building the template. For a wide range of cosmologies we have investigated in this paper, the best fit varies less than $1\%$ in the acoustic scale while the $1\sigma$ error is $\sim 4.7\%$. 

The angular distance scale we derive is $1.4 \sigma$ higher than the concordance $\LCDM$ model. When combined with three other BAO measurements from SDSS DR7 spectroscopic surveys at $z=0.2$ and 0.35 \citep{Percival10} and WiggleZ \citep{Blake11b} at $z \sim 0.6$, we find a tendency of cosmic distances measured using BAO to be larger than the concordance $\LCDM$ for $z \gtrsim 0.35$. Adding our measurement with these BAO measurements in the presence of WMAP7 prior therefore shifts the best fit $\Omega_m$ slightly larger than the concordance cosmology. 
 
\scr{In this paper, we have aimed at deriving a robust and conservative BAO information from the angular clustering of galaxies. We find that an accurate determination of the true redshift distribution of galaxies is crucial for a good photometric BAO measurement. Although the details of the method would and should vary for the conditions of different surveys, we hope that the approach described in this paper serves as a valuable reference for the analyses of future photometric BAO surveys.  }

\acknowledgements
We thank Chris Blake for providing the best fits and covariance matrix of $\DV/r_s$ measured using the WiggleZ data. We thank Patrick Mcdonald for helpful discussions.

Funding for SDSS-III has been provided by the Alfred P. Sloan Foundation, the Participating Institutions, the National Science Foundation, and the U.S. Department of Energy Office of Science. The SDSS-III web site is http://www.sdss3.org/.

SDSS-III is managed by the Astrophysical Research Consortium for the Participating Institutions of the SDSS-III Collaboration including the University of Arizona, the Brazilian Participation Group, Brookhaven National Laboratory, University of Cambridge, University of Florida, the French Participation Group, the German Participation Group, the Instituto de Astrofisica de Canarias, the Michigan State/Notre Dame/JINA Participation Group, Johns Hopkins University, Lawrence Berkeley National Laboratory, Max Planck Institute for Astrophysics, New Mexico State University, New York University, Ohio State University, Pennsylvania State University, University of Portsmouth, Princeton University, the Spanish Participation Group, University of Tokyo, University of Utah, Vanderbilt University, University of Virginia, University of Washington, and Yale University.

\end{document}